\documentclass[a4paper,fleqn]{cas-sc}
\usepackage{diagbox}
\usepackage[authoryear,longnamesfirst]{natbib}
\usepackage{wrapfig}
\usepackage[ruled,vlined]{algorithm2e}
\usepackage{algpseudocode} 
\usepackage{multirow}
\graphicspath{ {Figures/} }
\def\tsc#1{\csdef{#1}{\textsc{\lowercase{#1}}\xspace}}
\tsc{WGM}
\tsc{QE}
\tsc{EP}
\tsc{PMS}
\tsc{BEC}
\tsc{DE}

\begin{document}
\let\WriteBookmarks\relax
\def\floatpagepagefraction{1}
\def\textpagefraction{.001}
\shorttitle{Covid-19 Retractions Analysis}

\title [mode = title]{Unraveling Retraction Dynamics in COVID-19 Research: Patterns,
Reasons, and Implications}                      

\author[1]{Parul Khurana}

\author[2]{Ziya Uddin}

\author[2]{Kiran Sharma\corref{cor1}}
\ead{kiran.sharma@bmu.edu.in}
\cortext[cor1]{Corresponding author}
\address[1]{School of Computer Applications, Lovely Professional University, Phagwara, Punjab-144401, India}
\address[2]{School of Engineering and Technology, BML Munjal University, Gurugram, Haryana-122413, India}
%

\begin{abstract}
Amid the COVID-19 pandemic, while the world sought solutions, some scholars exploited the situation for personal gains through deceptive studies and manipulated data. This paper presents the extent of 400 retracted COVID-19 papers listed by the Retraction Watch database until the month of February 2024. The primary purpose of the research was to analyze journal quality and retractions trends. For all stakeholders involved, such as editors, relevant researchers, and policymakers, evaluating the journal's quality is crucial information since it could help them effectively stop such incidents and their negative effects in the future. The present research results imply that the one-fourth of publications were retracted within the first month of their publication, followed by an additional 6\% within six months of publication. One-third of the retractions originated from Q1 journals, with another significant portion coming from Q2 (29.8\%). A notable percentage of the retracted papers (23.2\%) lacked publishing impact, signifying their publication as conference papers or in journals not indexed by Scopus. An examination of the retraction reasons reveals that one-fourth of retractions were due to numerous causes, mostly in Q2 journals, and another quarter were due to data problems, with the majority happening in Q1 publications. Elsevier retracted 31\% of papers, with the majority published as Q1, followed by Springer (11.5\%), predominantly as Q2. On average, retractions from Q1 journals took 7.74 months, Q2 retractions took 10.44 months, and case reports had the longest duration at 12.3 months. Papers with specific reasons for retraction averaged over a year: fake-biased reviews took 15.08 months, multiple reasons took 14.64 months, and authorship issues took 12.63 months. Retracted papers were mainly associated with the USA, China, and India. In the USA, retractions were primarily from Q1 journals followed by no-impact publications; in China, it was Q1 followed by Q2, and in India, it was Q2 followed by no-impact publications. The study also examined author contributions, revealing that 69.3\% were male contributors, with females (30.7\%) mainly holding middle author positions. 

\end{abstract}


%
%
%
%


\maketitle

\section{Introduction}
With the proliferating speed of the COVID-19 pandemic, scientists, researchers, laboratories, and organizations examined the different aspects of the virus to determine the clinical, demographic, epidemiological, and radiological characteristics to assess the situation carefully~\citep{chamola2020comprehensive, khurana2023exploring}. The current state of research has undergone major changes due to the pressing need to comprehend the new disease and identify promising treatments~\citep{hirano2020covid}, as well as therapies~\citep{radecki2020impacts}. Many have taken advantage of the opportunity to enter the fields of infectious disease and healthcare research, and others already working in these areas have increased their intensity to trigger COVID-19-related research for the rapid dissemination of less stringently validated information~\citep{porter2020covid, park2021covid}.

This rapid outbreak and devastating severity adversely overburdened the academicians, scholars, and scientists, anticipating them to expedite their research for the flurry of unprecedentedly swift and quick publications~\citep{steen2011retractions, yeo2022sustained, fanelli2015misconduct}. In this phenomenon, the world witnessed an extraordinary global response, diverse aspects of the disease, and a large volume of scientific publications regarding the pandemic, associated virus, and its causative agent ~\citep{anderson2021academic, gupta2021impact}. The pandemic's rapid information production has sparked an informational epidemic, which may provide false information and sources of misinformation for scholarly communication~\citep{santos2021retractions}. Obviously, this pandemic situation has been stimulated in line with the steps taken by academic journals in the flow of the rapid peer review process and the concomitant surge in publications~\citep{borku2021most}.

However, there is a concern that such a high rate and intensity of additions to the literature may well be associated with chaotic changes in the research activities, compromise of scientific integrity, and advocacy of unusually high rates of retracted publications~\citep{london2020against, fang2012misconduct, yeo2021alarming}. Hence, the retraction of articles is another major challenge for science and society, where the reasons for retracting a publication are diverse and attributable to data~\citep{sharma2021team, chen2022characteristics}. Determining the nature of these retracted papers, the retraction of the publication may vary from an unintentional mistake to an intentional fraud of feverish activities~\citep{campos2019misconduct, delardas2022covid}. Retracted articles in medical science are a serious issue since they directly affect people's health~\citep{abhari2022twitter}. Hence, it is the equal responsibility of all those engaged in the publication of research, such as journals, editors, reviewers, writers, and publishers, to ensure the integrity of the process and of the information presented~\citep{malekpour2021scientific, shimray2023research}. Editors and editorial boards should review the grounds for retraction because this information can help stop similar situations and their harmful repercussions from happening again in the future~\citep{gholampour2022retracted}. Evidence, however, points to situations where universities have not put procedures in place to stop unethical research practices~\citep{lievore2021research}.

Although the retraction of the article is not a new phenomenon, there has been a lot of recent discussion about it. The scholarly literature can be corrected through retraction, which is gaining more attention as a means of ensuring that publications adhere to ethical standards and research rules~\citep{haunschild2021can, eldakar2023bibliometric}. Numerous studies on retracted papers have been conducted globally, with scholars examining this topic from various perspectives~\citep{teixeira2021silently,  sharma2021team, gai2021general, siva2023retracted, shimray2023research, anderson2021academic, cortegiani2021retracted}. Researchers have a consensus that retractions act as a mechanism for signalling errors; however, they have also had a number of negative effects, including contaminating citation networks, warping academic metrics, deceiving later research, hindering the advancement of science, undermining the reputation of science, squandering academic resources, and negatively impacting the careers of authors and co-authors~\citep{khademizadeh2023evolution, yuan2023research}.

Competent ethics committees that must continue to operate continuously oversee all research activities. While it is not appropriate to violate ethical standards, they can be modified to expedite the evaluation and endorsement of innovative methods. When it comes to ethics and ethical practices in clinical research, participants' autonomy must be respected, and beneficence, non-maleficence, and fairness must be guaranteed~\citep{xu2018retraction, el2021publications}. The ``Committee on Publication Ethics'' (COPE) has mandated that retraction announcements provide sufficient data about the retraction so that the paper's readers can take appropriate action in order to ensure transparent dissemination of the information~\citep{kleinert2009cope}.

Retraction Watch, which acts as a regulator for checking academic dishonesty, is maintaining a running list of retracted publications on COVID-19. Retraction Watch disseminated the curated list of 400 retracted publications on COVID-19. The rapid influx of publications was designated as articles, case reports, clinical studies, conference papers, reviews, and others ~\citep{van2020retracted, frampton2021inconsistent}.

\subsection{Research Objectives}

The objectives of the present study are
\begin{itemize}
\item Analyze the patterns and dynamics of retractions by studying the trends in retraction occurrences.
\item Investigate the primary reasons behind retractions to uncover the underlying cause of scientific misconduct.
\item Explore the trajectory of retracted articles in journals, considering their scientific impact and quartile rankings.
\item Explore geographic variations in retractions to identify countries with higher rates of retractions.
\item Examine the gender distribution among authors and their positions.

\end{itemize}

\section{Methodology}
\subsection{Data Collection}
The primary source of data on COVID-19 retracted publications is the ``Retraction Watch database''~\citep{watch2020retracted}, and data was extracted in the month of February 2024. A total of 400 publications were found, which were listed in the Retraction Watch Database under the ``Retracted coronavirus (COVID-19) papers" category. For all of these publications, the retraction status was verified manually from various sources, such as PubMed, Web of Science (WoS), Scopus, Google Scholar, and the publication journals. The fetched information in these 400 publications includes their title, author name, country name, affiliation, journal name, publisher name, publication subject, publication type, retraction reason, publication date, and retraction date as well.

\subsection{Author Filtration}

The next step is to filter the authors and their positions in the publications. To prepare the data for the analysis, we collected the author`s names with their position (First, Last, and Middle) in the publication. In total, we get a list of 2284 authors along with their respective positions in all 400 publications.

\subsection{Gender Identification}

To determine author gender, we used an online database named Gender API \url{(https://gender-api.com/en)}. This third-party gender API determines gender by name and country. This API has been used by many researchers in their work to examine the gender disparity in the authorship of academic documents\citep{andersen2020covid}. The gender extraction was performed based on the author's name and the country for the accuracy of the search results. Out of 2284 authors, the gender API returned the results for 1374 authors with an accuracy score ranging from 50 to 100. The accuracy score varies from 0 (not confirmed) to 100 (highly confirmed). Hence, we selected authors with accuracy scores above 60 for the analysis and 1271 authors out of 1374 were selected.

\begin{figure}[!h]
    \centering
    \includegraphics[width=\linewidth]{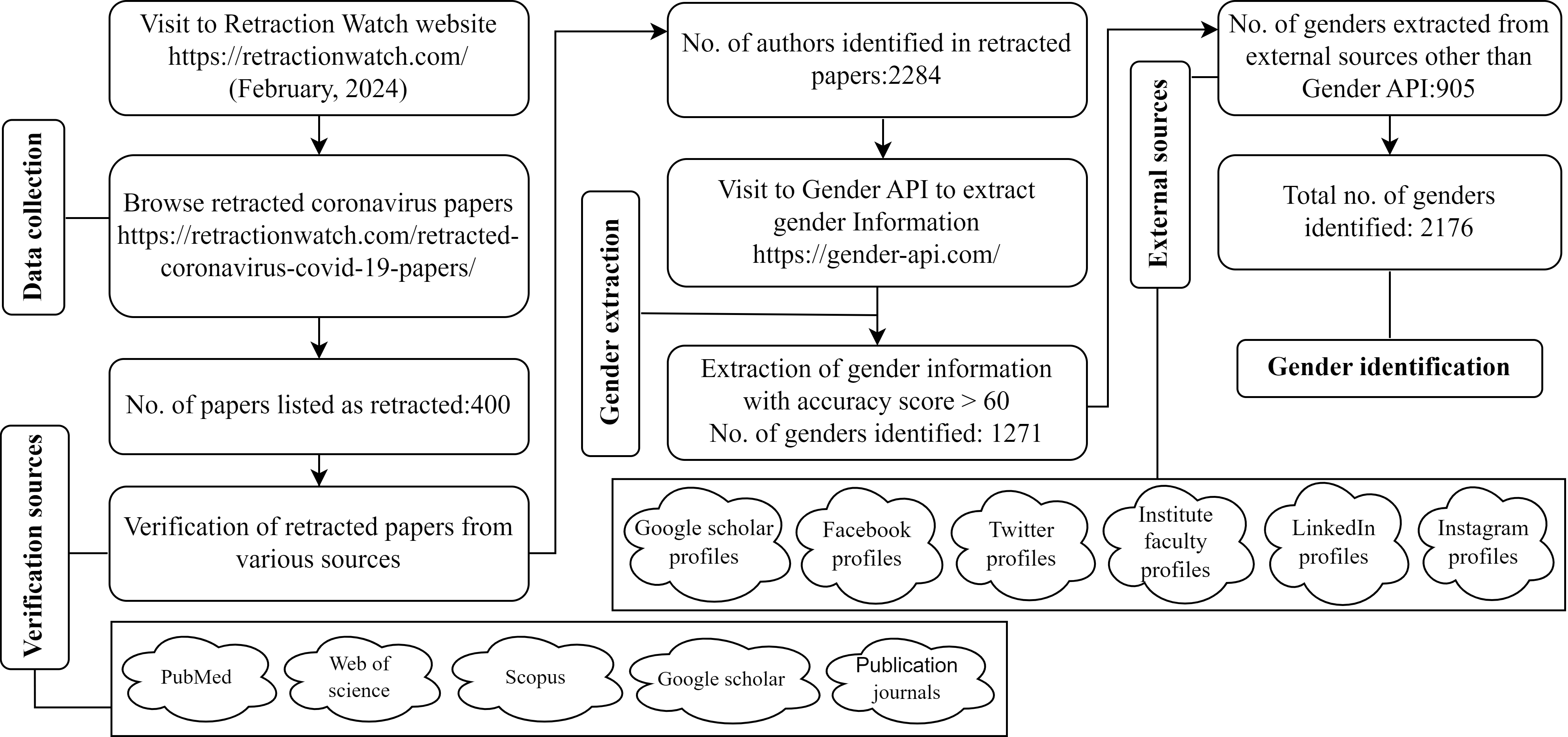}
\caption{Flowchart of data collection, verification and gender extraction.}
\label{fig:Fig1}   
\end{figure}

For rest of the 1013 authors, we checked Google scholar profiles, institute faculty profiles, Facebook profiles, Instagram profiles, Twitter profiles and LinkedIn profiles to find out the gender information of the authors manually. At the end, out of 1013, we were able to find gender information for 905 authors. Hence, 2176 out of 2284 authors were selected for the analysis. Figure~\ref{fig:Fig1} describes the complete process of data collection, verification and gender extraction.

\section{Results and Discussions}

\subsection{Retraction Timeline}

\begin{figure}[!h]
    \centering
      \includegraphics[width=\linewidth]{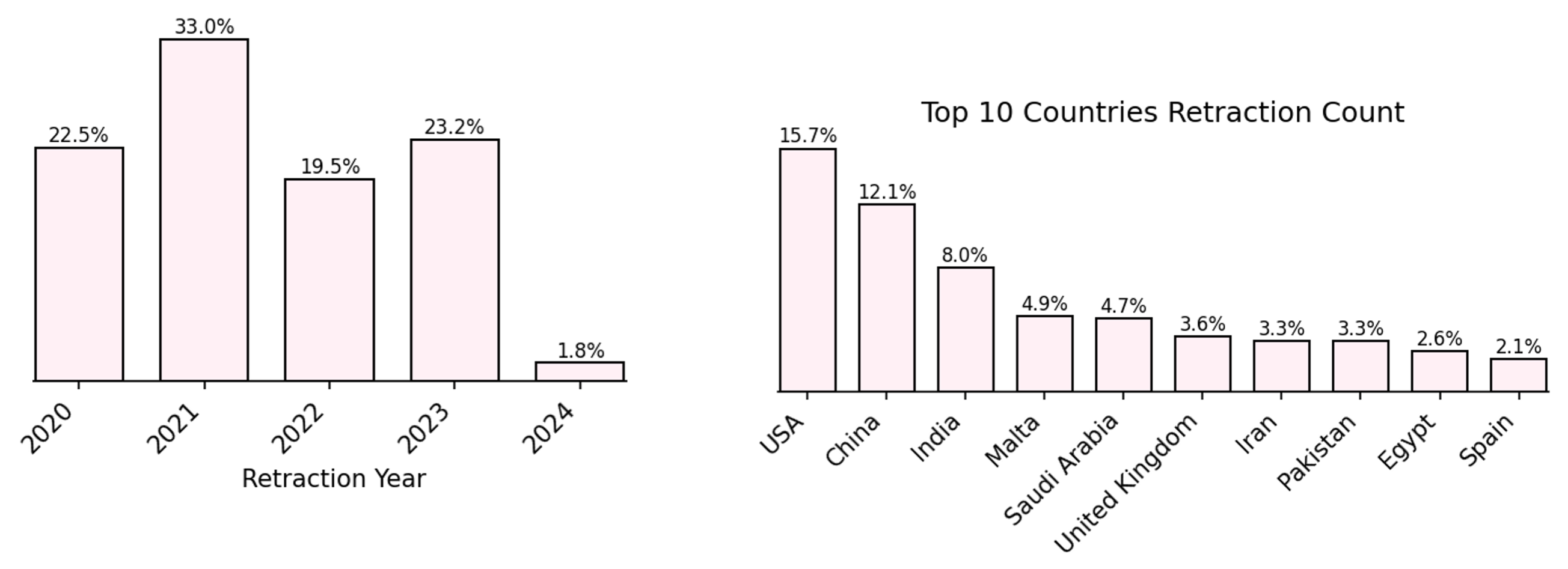}

\caption{(\textit{left}) Number of Covid-19 papers retracted from 2020-2024. (\textit{right}) Number of retracted papers of top 10 countries.}
\label{fig:Fig2}   
\end{figure}

Figure~\ref{fig:Fig2} illustrates the retraction timeline on the left, indicating that a significant portion of retractions occurred in the year 2021 (33\%) and in the year 2023 (23.2\%). On the right side of the figure, the top 10 countries with the highest number of retractions are highlighted, with the majority affiliated with the USA (15.7\%), followed by China (12.1\%), India (8\%), and so forth. Earlier studies also highlighted China and USA as the top contributor in retracted publications~\citep{yeo2021alarming}.

\subsection{Journal Impact}

\begin{figure}[!h]
    \centering
    \includegraphics[width=0.5\linewidth]{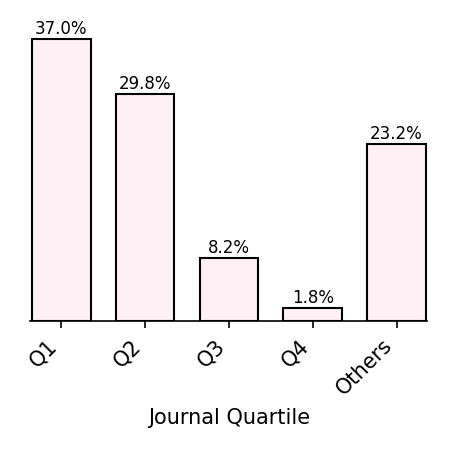}
\caption{Journal impact: Quartile wise proportion of the number of retracted publications. }
\label{fig:Fig3}   
\end{figure}

In addition to the impact factor, rankings of journals in each subject category are divided into quartiles by both Journal Citation Reports (JCR) (\url{https://clarivate.com}) and SCIMAGO Journal and Country Rank (SJR) (\url{https://www.scimagojr.com}). These quartiles rank the journals from highest to lowest based on their impact index.  Each subject category of journals is divided into four quartiles: Q1, Q2, Q3, Q4. Q1 is occupied by the top 25\% of journals in the list; Q2 is occupied by journals in the 25 to 50\% group; Q3 is occupied by journals in the 50 to 75\% group and Q4 is occupied by journals in the 75 to 100\% group.

The most prestigious journals within a subject area are those occupying the first quartile, Q1. Figure~\ref{fig:Fig3} depicts the distribution of papers (in \%) from each quartile. The highest proportion of retractions comes from Q1 (37\%), followed by Q2 (29.8\%). The third-largest category of retracted papers consists of those with no discernible impact (23.2\%), typically published in conferences, non-Scopus indexed journals, and occasionally as preprints. Further, the majority of retracted papers are articles (69.75\%), followed by reviews (12\%). Within the retracted articles, 26.25\% were originally published as Q1 papers, 21.75\% as Q2, and 14.75\% as no-impact papers. Similarly, the predominant retraction within the review category is attributed to Q2 type reviews (6.25\%). The corresponding numbers of retractions for each quartile and article type are detailed in Table~\ref{Table1}. The ``Others" category in article type includes Commentary, Editorial, Preprint, and Letter.


\begin{table}[!h]
\centering
\caption{Number of papers (in \%) as per the article type and journal impact (in Quartile). }

\label{Table1}
\begin{tabular}{|l|c|ccccc|c|}
\hline
\multirow{2}{*}{\textbf{Article Type}} &
  \multicolumn{1}{l|}{\multirow{2}{*}{\textbf{\begin{tabular}[c]{@{}l@{}}Total  \\ Publications\end{tabular}}}} &
  \multicolumn{5}{c|}{\textbf{Total Publications in each Quartile (\%)}} &
  \multicolumn{1}{l|}{\multirow{2}{*}{\textbf{\begin{tabular}[c]{@{}l@{}}Grand   \\ Total (\%)\end{tabular}}}} \\ \cline{3-7}
 &
  \multicolumn{1}{l|}{} &
  \multicolumn{1}{c|}{\textbf{Q1}} &
  \multicolumn{1}{c|}{\textbf{Q2}} &
  \multicolumn{1}{c|}{\textbf{Q3}} &
  \multicolumn{1}{c|}{\textbf{Q4}} &
  \textbf{Others} &
  \multicolumn{1}{l|}{} \\ \hline
Article &
  279 &
  \multicolumn{1}{c|}{26.25} &
  \multicolumn{1}{c|}{21.75} &
  \multicolumn{1}{c|}{5.50} &
  \multicolumn{1}{c|}{1.50} &
  14.75 &
  69.75 \\ \hline
Review &
  48 &
  \multicolumn{1}{c|}{3.50} &
  \multicolumn{1}{c|}{6.25} &
  \multicolumn{1}{c|}{1.50} &
  \multicolumn{1}{c|}{0} &
  0.75 &
  12.00 \\ \hline
Case Report &
  20 &
  \multicolumn{1}{c|}{1.25} &
  \multicolumn{1}{c|}{0} &
  \multicolumn{1}{c|}{0.75} &
  \multicolumn{1}{c|}{0} &
  3.00 &
  5.00 \\ \hline
Clinical Study &
  20 &
  \multicolumn{1}{c|}{2.75} &
  \multicolumn{1}{c|}{0.50} &
  \multicolumn{1}{c|}{0} &
  \multicolumn{1}{c|}{0.25} &
  1.50 &
  5.00 \\ \hline
Conference   Paper &
  14 &
  \multicolumn{1}{c|}{0.50} &
  \multicolumn{1}{c|}{0.50} &
  \multicolumn{1}{c|}{0} &
  \multicolumn{1}{c|}{0} &
  2.50 &
  3.50 \\ \hline
Others &
  19 &
  \multicolumn{1}{c|}{2.75} &
  \multicolumn{1}{c|}{0.75} &
  \multicolumn{1}{c|}{0.50} &
  \multicolumn{1}{c|}{0} &
  0.75 &
  4.75 \\ \hline
\textbf{Grand Total} &
  \textbf{400} &
  \multicolumn{1}{c|}{\textbf{37.00}} &
  \multicolumn{1}{c|}{\textbf{29.75}} &
  \multicolumn{1}{c|}{\textbf{8.25}} &
  \multicolumn{1}{c|}{\textbf{1.75}} &
  \textbf{23.25} &
  \textbf{100} \\ \hline
\end{tabular}
\end{table}

\subsection{Retraction Reasons}

Table~\ref{Table2} presents the reasons for retraction alongside the impact of the published papers, categorized into quartiles.  Reason of retraction are classified as~\citep{marcovitch2007misconduct, xu2021cross, campos2019misconduct}.
\begin{itemize}
\item \textit{Authorship issues and conflicts} encompass instances such as publishing without an author's approval, determining fictitious authors, disagreements between authors, false information provided by the corresponding author, and similar concerns.

\item \textit{Data-related concerns}  involve issues such as data manipulation, manipulation of figures, cases, or images, fabrication, falsification, and errors in methods.

\item \textit{Journal-related issues} encompass instances such as editorial duplication of content, uploading an incorrect manuscript or version, and duplication of articles.

\item  \textit{Plagiarism} involves the improper use of intellectual property belonging to individuals, encompassing articles, texts, study designs, tables, graphs, figures, and ideas. Self-plagiarism is also categorized under this heading.

\item \textit{Fake-biased peer review} includes fake or biased peer review processes and other issues related with the peer-review process.

\item \textit{Unethical research} involves instances such as the absence of approval from third parties, legal reasons or threats, lack of approval from companies or institutions, lack of Institutional Review Board (IRB) or Institutional Animal Care and Use Committee (IACUC) approval, and breaches of author policies.

\item \textit{Multiple reasons} refer to papers with more than one cause for retraction. These reasons may include issues such as authorship conflicts, data-related concerns, journal-related problems, plagiarism, biased peer reviews, unethical research, conflicts of interest, etc.

\item \textit{Unknown} category comprises instances where the reason for retraction is either unspecified or unclear.

\end{itemize}

The primary causes of retractions, along with the corresponding number of retracted papers, include multiple reasons (25.5\%),  data-related concerns (23.25\%), and journal-related issues (13\%) as shown in Table~\ref{Table2}. As depicted in Figure~\ref{fig:Fig3}, the majority of retractions stem from Q1 journals (37\%), among which 10.25\% were retracted due to data integrity issues, 6.75\% due to multiple reasons, 5.5\% due to unknown reasons, and 5.25\% due to journal-related issues. The second-largest proportion of retractions originates from Q2 journals (29.75\%), with 12\% attributed to multiple reasons and 8.5\% to unknown reasons.The third-highest proportion of retractions (23.25\%) pertains to papers with no discernible impact, typically published in conferences, preprints, or non-Scopus indexed journals. The primary reasons for retraction in this category include data-related issues (8.5\%) and multiple reasons (5.25\%).

\begin{table}[!h]
\centering
\caption{Number of papers (in \%) as per retraction reasons and journal impact (in Quartile). }
\label{Table2}
\begin{tabular}{|l|c|ccccc|c|}
\hline
\multirow{2}{*}{\textbf{Retraction   Reasons}} &
  \multicolumn{1}{l|}{\multirow{2}{*}{\textbf{\begin{tabular}[c]{@{}l@{}}Total   \\ Publications\end{tabular}}}} &
  \multicolumn{5}{c|}{\textbf{Total Publications in each Quartile (\%)}} &
  \multicolumn{1}{l|}{\multirow{2}{*}{\textbf{\begin{tabular}[c]{@{}l@{}}Grand \\ Total (\%)\end{tabular}}}} \\ \cline{3-7}
 &
  \multicolumn{1}{l|}{} &
  \multicolumn{1}{c|}{\textbf{Q1}} &
  \multicolumn{1}{c|}{\textbf{Q2}} &
  \multicolumn{1}{c|}{\textbf{Q3}} &
  \multicolumn{1}{c|}{\textbf{Q4}} &
  \textbf{Others} &
  \multicolumn{1}{l|}{} \\ \hline
Authorship Issues and Conflicts &
  8 &
  \multicolumn{1}{c|}{1.25} &
  \multicolumn{1}{c|}{0.25} &
  \multicolumn{1}{c|}{0} &
  \multicolumn{1}{c|}{0} &
  0.50 &
  2.00 \\ \hline
Data-related Concerns &
  93 &
  \multicolumn{1}{c|}{10.25} &
  \multicolumn{1}{c|}{3.25} &
  \multicolumn{1}{c|}{1.00} &
  \multicolumn{1}{c|}{0.25} &
  8.50 &
  23.25 \\ \hline
Journal-related Issues &
  52 &
  \multicolumn{1}{c|}{5.25} &
  \multicolumn{1}{c|}{3.00} &
  \multicolumn{1}{c|}{3.00} &
  \multicolumn{1}{c|}{0.50} &
  1.25 &
  13.00 \\ \hline
Multiple Reasons &
  102 &
  \multicolumn{1}{c|}{6.75} &
  \multicolumn{1}{c|}{12.00} &
  \multicolumn{1}{c|}{1.00} &
  \multicolumn{1}{c|}{0.50} &
  5.25 &
  25.50 \\ \hline
Plagiarism &
  19 &
  \multicolumn{1}{c|}{2.75} &
  \multicolumn{1}{c|}{1.00} &
  \multicolumn{1}{c|}{0.50} &
  \multicolumn{1}{c|}{0} &
  0.50 &
  4.75 \\ \hline
Fake-biased Peer Review &
  24 &
  \multicolumn{1}{c|}{3.25} &
  \multicolumn{1}{c|}{1.25} &
  \multicolumn{1}{c|}{0.25} &
  \multicolumn{1}{c|}{0} &
  1.25 &
  6.00 \\ \hline
Unethical Research &
  18 &
  \multicolumn{1}{c|}{2.00} &
  \multicolumn{1}{c|}{0.50} &
  \multicolumn{1}{c|}{0.25} &
  \multicolumn{1}{c|}{0.25} &
  1.50 &
  4.50 \\ \hline
Unknown &
  84 &
  \multicolumn{1}{c|}{5.50} &
  \multicolumn{1}{c|}{8.50} &
  \multicolumn{1}{c|}{2.25} &
  \multicolumn{1}{c|}{0.25} &
  4.50 &
  21.00 \\ \hline
\textbf{Grand Total} &
  \textbf{400} &
  \multicolumn{1}{c|}{\textbf{37.00}} &
  \multicolumn{1}{c|}{\textbf{29.75}} &
  \multicolumn{1}{c|}{\textbf{8.25}} &
  \multicolumn{1}{c|}{\textbf{1.75}} &
  \textbf{23.25} &
  \textbf{100} \\ \hline
\end{tabular}
\end{table}


Table~\ref{Table_Publisher} presents the retraction statistics based on publishing quality for the top 10 publishing houses. The highest retraction rate is observed for Elsevier (31\%), followed by Springer (11.5\%), and others. Specifically, the retraction rates for Cold Spring Harbor Laboratory Press and Cureus are 7.75\% and 4.25\%, respectively. Notably, all retractions from these publishers are associated with non-impact (no journal quartile) papers. In contrast, retractions from Elsevier, Springer, Hindawi, and SAGE Publications span across all categories of journals. For instance, a significant portion of retractions from Elsevier is from Q1 (41.94\%), while Springer has a majority from Q2 (54.35\%). Hindawi predominantly experiences retractions from Q2 (76.92\%), and SAGE Publications mainly from Q1 (50\%). Furthermore, the majority of retractions from Wiley are from Q2 (75\%), while retractions from Oxford Academic are predominantly from Q1 (87.5\%).
\begin{table}[!h]
\centering
\caption{Number of retracted papers (in \%) from top 10 publishing houses and journal impact (in Quartile). }
\label{Table_Publisher}
\begin{tabular}{|l|cc|ccccc|}
\hline
\multirow{2}{*}{\textbf{Publisher}} &
  \multicolumn{2}{l|}{\textbf{Total Retractions}} &
  \multicolumn{5}{c|}{\textbf{Journal Quartile}} \\ \cline{2-8} 
 &
  \multicolumn{1}{c|}{\textbf{Count}} &
  \textbf{\%age} &
  \multicolumn{1}{c|}{\textbf{Q1}} &
  \multicolumn{1}{c|}{\textbf{Q2}} &
  \multicolumn{1}{c|}{\textbf{Q3}} &
  \multicolumn{1}{c|}{\textbf{Q4}} &
  \textbf{Others} \\ \hline
Elsevier &
  \multicolumn{1}{c|}{124} &
  31.00 &
  \multicolumn{1}{c|}{41.94} &
  \multicolumn{1}{c|}{35.48} &
  \multicolumn{1}{c|}{12.10} &
  \multicolumn{1}{c|}{0.81} &
  9.68 \\ \hline
Springer &
  \multicolumn{1}{c|}{46} &
  11.50 &
  \multicolumn{1}{c|}{36.96} &
  \multicolumn{1}{c|}{54.35} &
  \multicolumn{1}{c|}{2.17} &
  \multicolumn{1}{c|}{2.17} &
  4.35 \\ \hline
Cold Spring Harbor Laboratory Press &
  \multicolumn{1}{c|}{31} &
  7.75 &
  \multicolumn{1}{c|}{0} &
  \multicolumn{1}{c|}{0} &
  \multicolumn{1}{c|}{0} &
  \multicolumn{1}{c|}{0} &
  100 \\ \hline
Wiley &
  \multicolumn{1}{c|}{24} &
  6.00 &
  \multicolumn{1}{c|}{20.83} &
  \multicolumn{1}{c|}{75.00} &
  \multicolumn{1}{c|}{0} &
  \multicolumn{1}{c|}{0} &
  4.17 \\ \hline
Cureus &
  \multicolumn{1}{c|}{17} &
  4.25 &
  \multicolumn{1}{c|}{0} &
  \multicolumn{1}{c|}{0} &
  \multicolumn{1}{c|}{0} &
  \multicolumn{1}{c|}{0} &
  100 \\ \hline
Frontiers &
  \multicolumn{1}{c|}{16} &
  4.00 &
  \multicolumn{1}{c|}{75.00} &
  \multicolumn{1}{c|}{6.25} &
  \multicolumn{1}{c|}{0} &
  \multicolumn{1}{c|}{0} &
  18.75 \\ \hline
Taylor and Francis &
  \multicolumn{1}{c|}{15} &
  3.75 &
  \multicolumn{1}{c|}{80.00} &
  \multicolumn{1}{c|}{20.00} &
  \multicolumn{1}{c|}{0} &
  \multicolumn{1}{c|}{0} &
  0 \\ \hline
Hindawi &
  \multicolumn{1}{c|}{13} &
  3.25 &
  \multicolumn{1}{c|}{15.38} &
  \multicolumn{1}{c|}{76.92} &
  \multicolumn{1}{c|}{7.69} &
  \multicolumn{1}{c|}{0} &
  0 \\ \hline
SAGE Publications &
  \multicolumn{1}{c|}{12} &
  3.00 &
  \multicolumn{1}{c|}{50.00} &
  \multicolumn{1}{c|}{8.33} &
  \multicolumn{1}{c|}{16.67} &
  \multicolumn{1}{c|}{25.00} &
  0 \\ \hline
Oxford Academic &
  \multicolumn{1}{c|}{8} &
  2.00 &
  \multicolumn{1}{c|}{87.50} &
  \multicolumn{1}{c|}{12.50} &
  \multicolumn{1}{c|}{0} &
  \multicolumn{1}{c|}{0} &
  0 \\ \hline
\end{tabular}
\end{table}
Similarly, Table~\ref{Table4} shows the distribution of retracted papers based on article type and the reason for retraction. As previously discussed, the majority of retractions involve articles (69.5\%), with 19.5\% retracted due to multiple reasons, 16.5\% due to data-related issues, and 15\% due to unknown reasons. The second-highest category of retracted papers is reviews (12\%), with the predominant reasons for retraction being unknown (4.25\%) and multiple reasons (2.75\%). The retraction rates in the remaining categories are relatively low.


\begin{table}[!h]
\caption{Number of retracted papers (in \%) as per article type and retraction reasons. }
\label{Table4}
\begin{tabular}{|l|c|cccccc|c|}
\hline
\multirow{2}{*}{\textbf{Article Type}} &
  \multicolumn{1}{l|}{\multirow{2}{*}{\textbf{\begin{tabular}[c]{@{}l@{}}Total\\ Publications\end{tabular}}}} &
  \multicolumn{6}{c|}{\textbf{Total Publications in each Category (\%)}} &
  \multicolumn{1}{l|}{\multirow{2}{*}{\textbf{\begin{tabular}[c]{@{}l@{}}Grand \\ Total\\  (\%)\end{tabular}}}} \\ \cline{3-8}
 &
  \multicolumn{1}{l|}{} &
  \multicolumn{1}{c|}{\textbf{Article}} &
  \multicolumn{1}{c|}{\textbf{Review}} &
  \multicolumn{1}{c|}{\textbf{\begin{tabular}[c]{@{}c@{}}Case \\ Report\end{tabular}}} &
  \multicolumn{1}{c|}{\textbf{\begin{tabular}[c]{@{}c@{}}Clinical\\ Study\end{tabular}}} &
  \multicolumn{1}{c|}{\textbf{\begin{tabular}[c]{@{}c@{}}Conference \\ Paper\end{tabular}}} &
  \textbf{Other} &
  \multicolumn{1}{l|}{} \\ \hline
\begin{tabular}[c]{@{}l@{}}Authorship Issues \\ and Conflicts\end{tabular} &
  8 &
  \multicolumn{1}{c|}{1.50} &
  \multicolumn{1}{c|}{0.25} &
  \multicolumn{1}{c|}{0} &
  \multicolumn{1}{c|}{0} &
  \multicolumn{1}{c|}{0} &
  0.25 &
  2.00 \\ \hline
Data-related Concerns &
  93 &
  \multicolumn{1}{c|}{16.50} &
  \multicolumn{1}{c|}{1.25} &
  \multicolumn{1}{c|}{1.50} &
  \multicolumn{1}{c|}{2.25} &
  \multicolumn{1}{c|}{1.00} &
  0.75 &
  23.25 \\ \hline
Journal-related Issues &
  52 &
  \multicolumn{1}{c|}{7.00} &
  \multicolumn{1}{c|}{2.00} &
  \multicolumn{1}{c|}{0.75} &
  \multicolumn{1}{c|}{0.50} &
  \multicolumn{1}{c|}{0.75} &
  2.00 &
  13.00 \\ \hline
Multiple Reasons &
  102 &
  \multicolumn{1}{c|}{19.00} &
  \multicolumn{1}{c|}{2.75} &
  \multicolumn{1}{c|}{1.50} &
  \multicolumn{1}{c|}{1.00} &
  \multicolumn{1}{c|}{0.50} &
  0.75 &
  25.50 \\ \hline
Plagiarism &
  19 &
  \multicolumn{1}{c|}{3.25} &
  \multicolumn{1}{c|}{1.00} &
  \multicolumn{1}{c|}{0.25} &
  \multicolumn{1}{c|}{0} &
  \multicolumn{1}{c|}{0.25} &
  0 &
  4.75 \\ \hline
Fake-biased Peer review &
  24 &
  \multicolumn{1}{c|}{4.50} &
  \multicolumn{1}{c|}{0.50} &
  \multicolumn{1}{c|}{0.25} &
  \multicolumn{1}{c|}{0} &
  \multicolumn{1}{c|}{0.75} &
  0 &
  6.00 \\ \hline
Unethical Research &
  18 &
  \multicolumn{1}{c|}{3.00} &
  \multicolumn{1}{c|}{0} &
  \multicolumn{1}{c|}{0.75} &
  \multicolumn{1}{c|}{0.50} &
  \multicolumn{1}{c|}{0} &
  0.25 &
  4.50 \\ \hline
Unknown &
  84 &
  \multicolumn{1}{c|}{15.00} &
  \multicolumn{1}{c|}{4.25} &
  \multicolumn{1}{c|}{0} &
  \multicolumn{1}{c|}{0.75} &
  \multicolumn{1}{c|}{0.25} &
  0.75 &
  21.00 \\ \hline
\textbf{Grand Total} &
  \textbf{400} &
  \multicolumn{1}{c|}{\textbf{69.75}} &
  \multicolumn{1}{c|}{\textbf{12.00}} &
  \multicolumn{1}{c|}{\textbf{5.00}} &
  \multicolumn{1}{c|}{\textbf{5.00}} &
  \multicolumn{1}{c|}{\textbf{3.50}} &
  \textbf{4.75} &
  \textbf{100} \\ \hline
\end{tabular}
\end{table}

\subsection{Time to Retraction}

\begin{figure}[!h]
    \centering
    \includegraphics[width=0.85\linewidth]{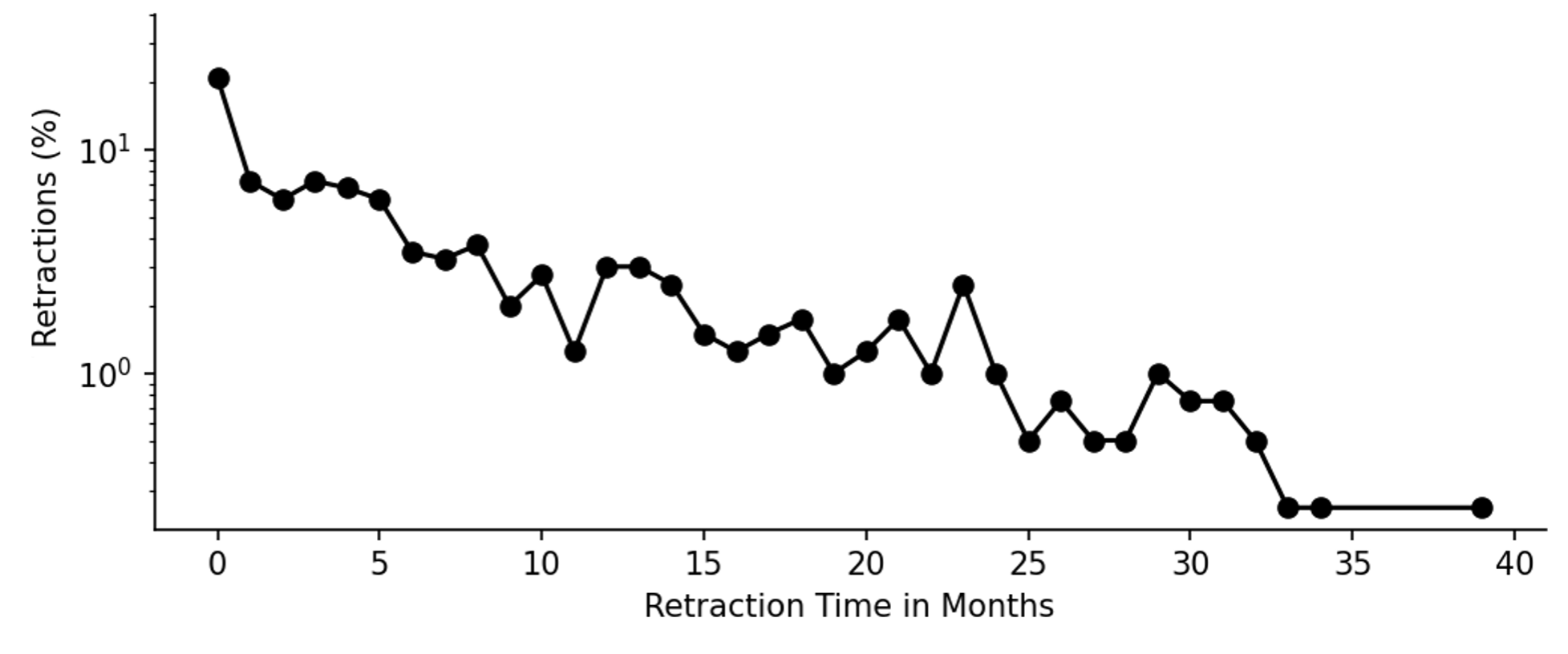}
\caption{Time to retraction (in months) versus number of retractions. }
\label{fig:Fig4}   
\end{figure}

Figure~\ref{fig:Fig4} displays the retraction time (in months) and the corresponding number of retractions during that timeframe. Approximately 21\% of papers were retracted within one month of publication, followed by 7.25\% after one month, 6.25\% after two months, 7.25\% after three months, and so forth. Around 3\% of papers were retracted after one year, 1\% after two years, and 0.25\% after three years of publication.
Primarily, a significant majority of retractions occurred within the first six months of publication. Retractions in the Q1 category were mostly addressed within this six-month timeframe. The majority of papers retracted after a year belong to the Q2 and Q3 categories.

Table~\ref{Table3} presents the average time taken by papers, categorized by the publishing journal impact, article type, and retraction reason. On average, the time taken for each retraction is 8.01 months, with the highest average time reported for Q4 papers (12.56), followed by Q2 (10.44), and Q1 (7.74). Despite the higher number of retracted papers in the Q1 category, the average time for retraction is comparatively less (7.74) than other categories.Similarly, the maximum time was taken by case reports (12.3), followed by conference papers (11.36), and clinical studies (8.55). Although the number of retractions from these categories is lower compared to articles and reviews. The retraction time is shorter for articles and reviews compared to conference papers, clinical studies, and case reports.

Furthermore, in terms of retraction reasons, the maximum time was taken by papers associated with fake-biased peer review (15.08), followed by the multiple reasons category (14.64). The majority of retractions are attributed to multiple reasons, followed by data-related concerns; however, retracted papers due to data-related issues took less time (6.75) compared to those retracted for multiple reasons. The third-highest retracted category is unknown reasons, with a lesser retraction time (2.51).

\begin{table}[!h]
\centering
\caption{The average duration for paper retractions (measured in months) varies among journals of different quartiles, various article types, and different retraction reasons.}

\label{Table3}
\begin{tabular}{|l|c|l|l|c|l|l|c|}
\hline
\begin{tabular}[c]{@{}l@{}}Journal   \\ Impact\end{tabular} &
  \multicolumn{1}{l|}{\begin{tabular}[c]{@{}l@{}}Average \\ Retraction\\ Time \\ (in months)\end{tabular}} &
   &
  Article Type &
  \multicolumn{1}{l|}{\begin{tabular}[c]{@{}l@{}}Average \\ Retraction\\ Time \\ (in months)\end{tabular}} &
   &
  Retraction Reasons &
  \multicolumn{1}{l|}{\begin{tabular}[c]{@{}l@{}}Average \\ Retraction\\ Time \\ (in months)\end{tabular}} \\ \hline
Q1     & 7.74                  &  & Article          & 7.92                  &  & Authorship Issues and Conflicts & 12.63 \\ \hline
Q2     & 10.44                 &  & Review           & 7.15                  &  & Data-related Concerns           & 6.75  \\ \hline
Q3     & 3.21                  &  & Case Report      & 12.3                  &  & Journal-related Issues          & 3.13  \\ \hline
Q4     & 12.57                 &  & Clinical Study   & 8.55                  &  & Multiple Reasons                & 14.64 \\ \hline
Others & 6.69                  &  & Conference Paper & 11.36                 &  & Plagiarism                      & 7.74  \\ \hline
 - & 	-			&  & Others            & 4                     &  & Fake-biased Peer Review         & 15.08 \\ \hline
 - & 	-	 &  &                 - & 	-			 &  & Unethical Research              & 5.5   \\ \hline
 - & 	-			 &  &                   - & 	-			 &  & Unknown                         & 2.51  \\ \hline
\end{tabular}
\end{table}

\subsection{Country Collaboration Network}
\begin{figure}[!h]
    \centering
    \includegraphics[width=\linewidth]{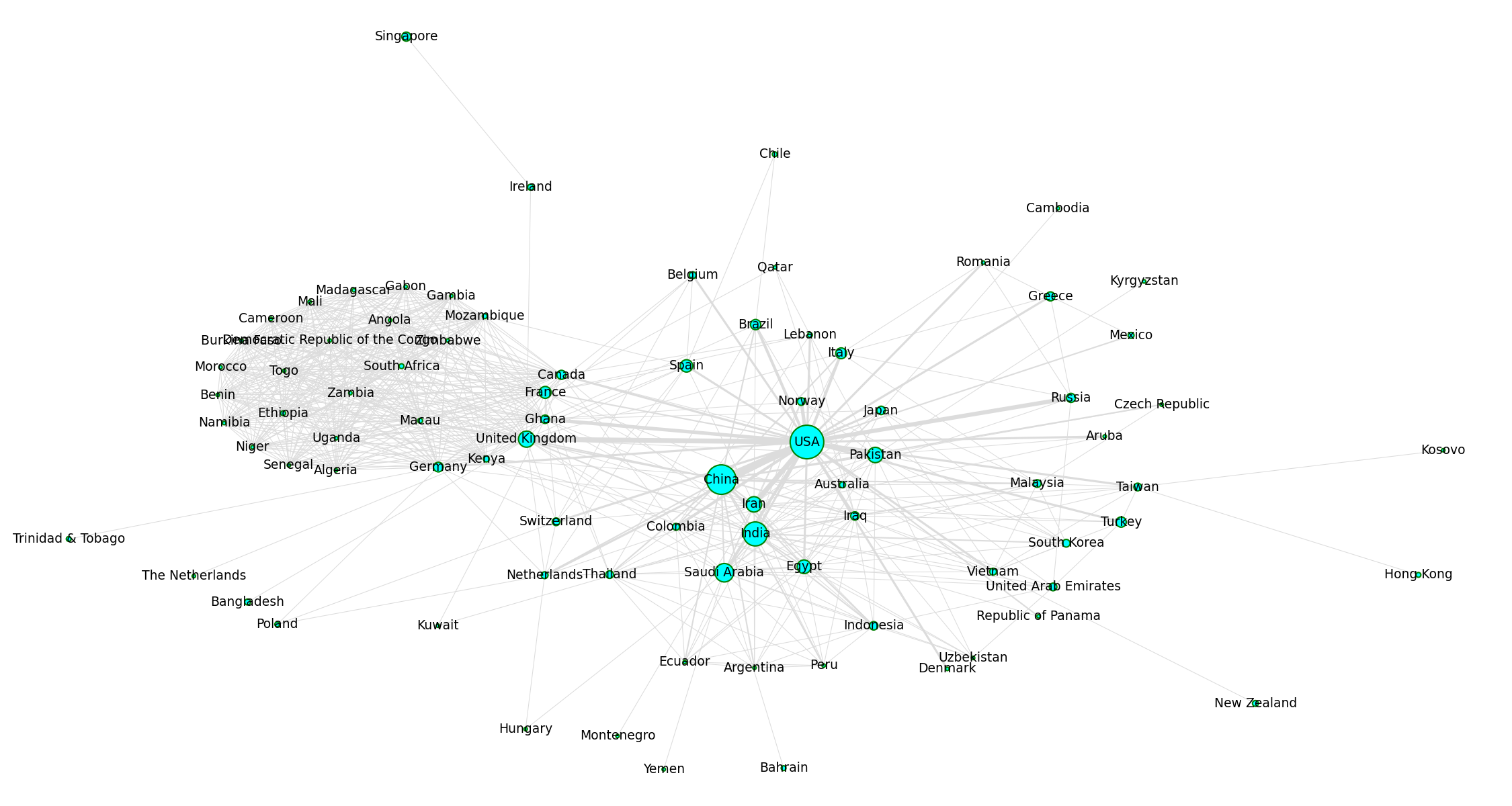}
\caption{Country Collaboration network of retracted publications.}
\label{fig:Fig5}   
\end{figure}

Figure~\ref{fig:Fig5} illustrates the collaboration network among countries in retracted papers. Papers retracted due to Covid-19 involved authors from 99 different countries, fostering intercountry collaboration among 85 of them (represented by the number of nodes). The network consists of 633 edges, where the edge width signifies the strength of collaboration (the number of times authors from the same country collaborated). The node size corresponds to the number of papers affiliated with each country.

Examining Fig~\ref{fig:Fig2} (\textit{right}), it is evident that the highest number of retractions is attributed to the USA, followed by China and India. Interestingly, Malta ranks fourth in retracted papers, yet all retractions from Malta are associated with intra-country collaboration.
The grouping of nodes on the left side of the network, including countries such as Zambia, Kenya, Ghana, etc., represents a singular paper that involves collaboration among 27 different countries. 
The leading collaborators among the retracted papers are the USA and China, demonstrating the most robust collaboration, followed by India and Saudi Arabia, as well as the USA and the United Kingdom, the USA and Iran, and so on.

\subsection{Inter-country and Intra-country Collaboration Analysis}

\begin{figure}[!h]
    \centering
     \includegraphics[width=0.6\linewidth]{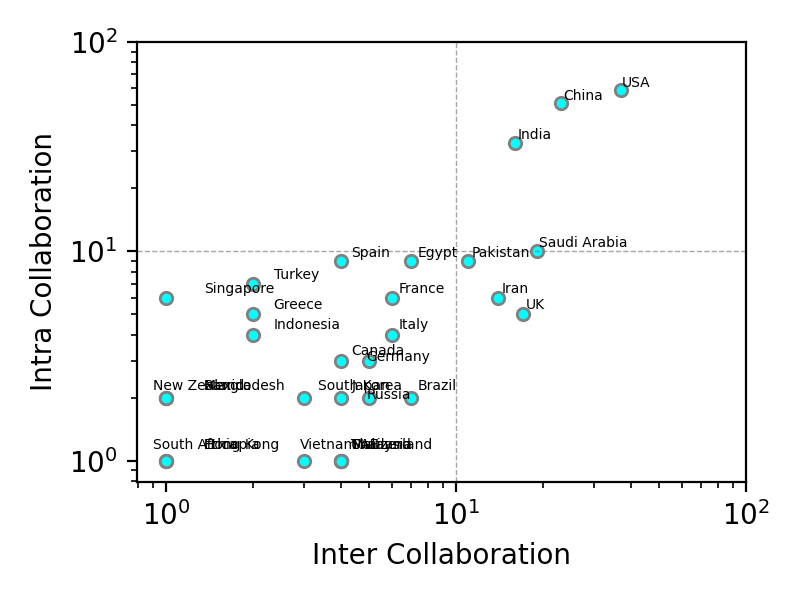}
\caption{Scatter plot representing the inter- country and intra-country collaboration.}
\label{fig:Fig6}   
\end{figure}

Figure~\ref{fig:Fig6} depicts a comparison between collaborations at the inter-country and intra-country levels. Countries lacking collaboration in either category are excluded from the figure. Notably, the USA, China, and India form a distinct cluster in the illustration. In the case of retractions from the USA, 38.54\% papers were associated with international collaboration, while 61.45\% involved national collaboration. Similarly, China experienced 31.08\% retractions with international collaboration and 68.91\% with national collaboration. For India, 32.65\% of retractions involved international collaborators, compared to 67.34\% with national collaboration. The pattern of retractions concerning collaborations is remarkably similar in the top three retracted countries. In all three cases, a notable trend emerges where a significant portion of retractions is attributed to national collaboration rather than international collaboration.Conversely, Saudi Arabia, the United Kingdom, Iran, and Pakistan witnessed a higher proportion of retractions associated with international collaboration compared to national collaboration. 

\subsubsection{Patterns of Retraction- Top Three Countries}

\begin{figure}[!h]
    \centering
   \includegraphics[width=\linewidth]{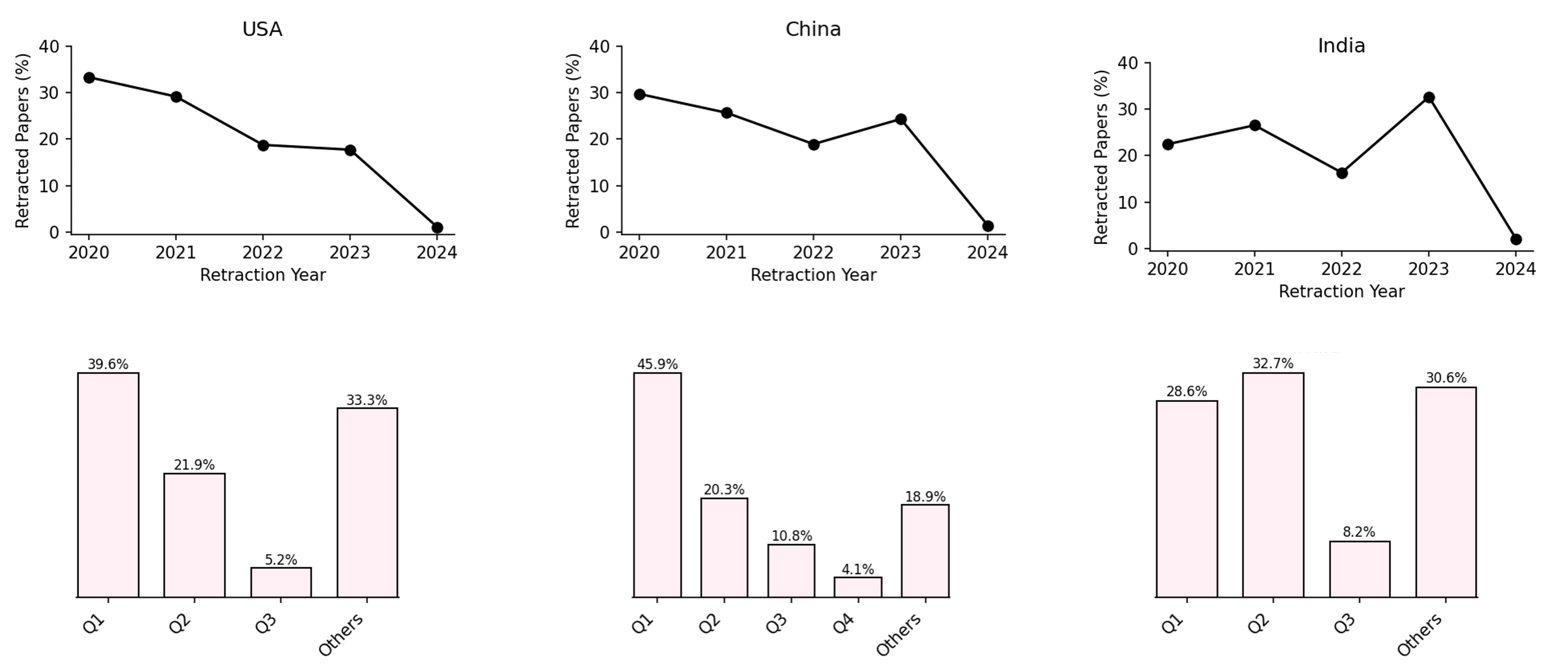}

\caption{Number of papers retracted per year and journal impact of retracted papers. (\textit{left}) USA, (\textit{middle}) China, and (\textit{right}) India;}
\label{fig:Fig7}   
\end{figure}

We conducted an in-depth analysis of retraction patterns in papers originating from the top three countries: the USA, China, and India (see Figure~\ref{fig:Fig6}). Figure~\ref{fig:Fig7} (top) illustrates the retraction rate trends for all three countries. Interestingly, the retraction rate in the USA has consistently decreased over the years: 2020 (33.3\%), 2021 (29.2\%), 2022 (18.7\%), and 2023 (17.7\%). In contrast, China shows a declining trend in 2020 (29.7\%), 2021 (25.6\%), and 2022 (18.9\%), but experiences a subsequent increase in 2023 (24.3\%). India exhibits a distinct pattern with a rate of 22.4\% in 2020, followed by a rise in 2021 (26.5\%), a decline in 2022 (16.3\%), and another increase in 2023 (32.6\%).

On the other hand, figure~\ref{fig:Fig7} (bottom) depicts the total retractions corresponding to each country based on their publishing impact (journal quartile). 
In the USA, major retractions occurred in Q1 publishing journals (39.6\%), followed by 21.9\% in Q2. The second-highest retracted papers were from no-impact journals, constituting 33.3\%. In China, the majority of retractions were in the Q1 category (45.9\%), followed by Q2 (20.3\%). Q2 retractions were the second-highest in China. In India, the primary retractions were from Q2 (32.7\%), followed by Q1 (28.6\%). The second-highest retraction in India was from non-impact papers (30.6\%). Interestingly, Malta ranked as the fourth highest in retractions, with all retractions originating from Q2 publications. In comparison, Saudi Arabia also listed 55.6\% of retractions from Q2 publications. 
In contrast to the USA and China, India exhibits a nearly identical proportion of retractions across Q1, Q2, and no-impact journals. 


\subsection{Author's Proportion and Position Analysis}

\begin{figure}[!h]
    \centering
   \includegraphics[width=0.75\linewidth]{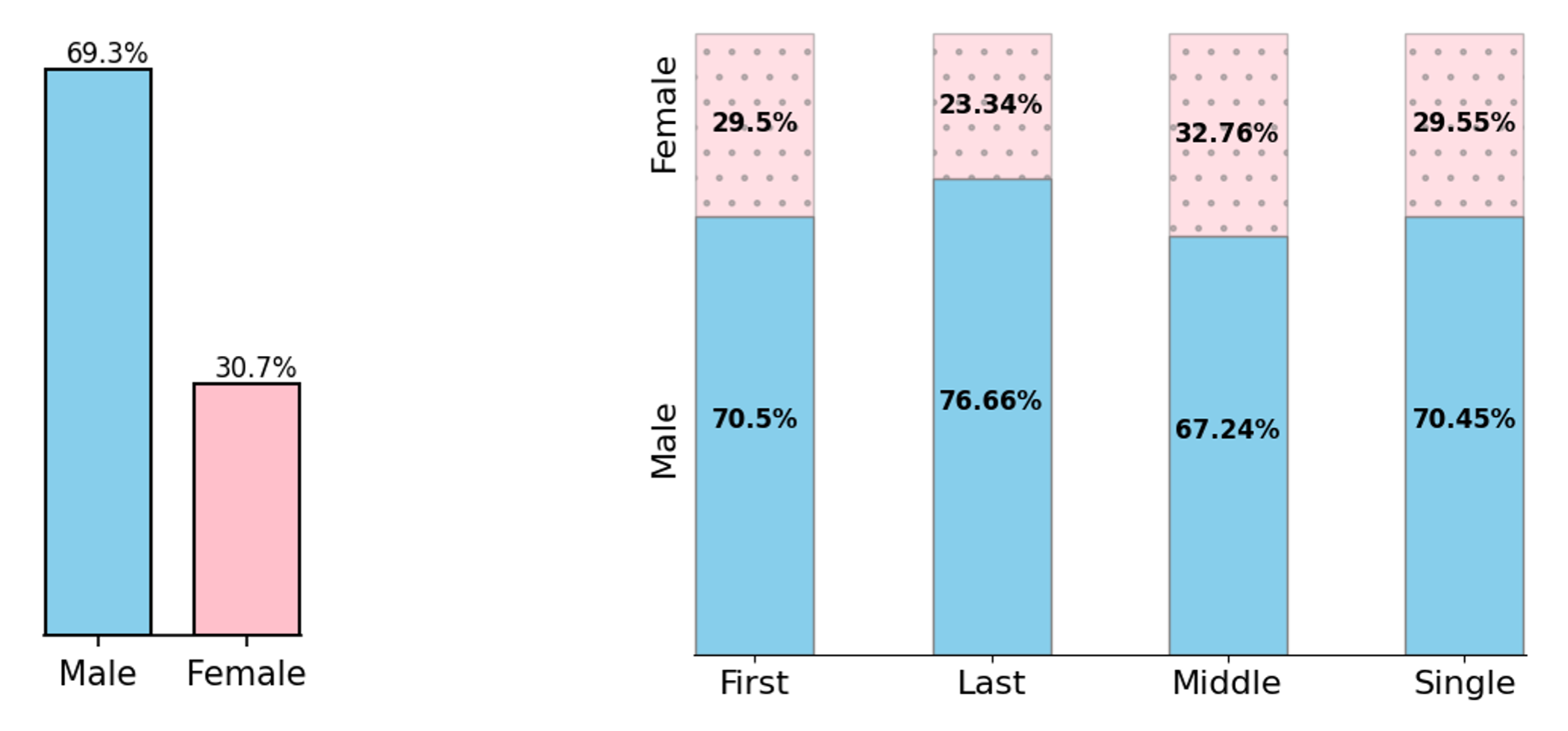}

\caption{(\textit{left}) Gender proportion in retracted publications. (\textit{right}) Proportion of female authorship position as first authors, last authors, middle authors,  and single author in Covid-19 retracted publications. }
\label{fig:Fig8}   
\end{figure}

Figure~\ref{fig:Fig8} (left) illustrates the distribution of genders in retracted publications. Among 2176 authors, 69.3\% of retracted papers are authored by males, while 30.7\% are authored by females. Previous studies have indicated a connection between gender and authorship positions~\citep{nielsen2017one, penner2015gender}. Additionally, earlier research has pointed out that COVID-19 medical papers exhibit a lower-than-expected representation of women as first authors~\citep{andersen2020covid}.
Figure~\ref{fig:Fig8} (\textit{right}) shows the proportion of female authors in retracted papers at different authorship position. Majority of the female contribution is spotted at middle position (32.76\%) as compared to first (29.5\%) and last (23.34\%) position. Single female author retracted paper proportion is 29.5\%. 

A notable majority, amounting to 61.5\%, of papers retracted with a sole female author were initially published in Q1 journals. The primary reasons for these retractions were related to data integrity issues, and these papers were frequently categorized as articles. When examining the position of females as the first author in retracted papers, the majority of retractions occurred in Q1 journals (42.86\%), followed by Q2 journals (31.63\%). In the middle authorship position, a significant portion of retractions were from Q1 journals (42.12\%), followed by journals with no impact (25.93\%). Similarly, in the last authorship position, Q1 retracted papers constituted 39.24\%, with Q2 closely following at 34.18\%. 
On average, 41.8\% of papers with contributions from female authors faced retractions from Q1 Journals, while 27\% were retracted from Q2. Additionally,  on average, 28.6\% of retractions were attributed to data integrity issues, and 27.6\% had multiple reasons when female authors were co-authors. Similarly, on average,, 65.1\% of retracted papers, where female authors made contributions, were initially published as articles, and 12.7\% as clinical studies. Moreover, among the papers retracted from the top three countries, 33.3\% of female authors are affiliated with the USA, 30.3\% from China, and 22.9\% from India.

\begin{figure}[!h]
    \centering
   \includegraphics[width=0.47\linewidth]{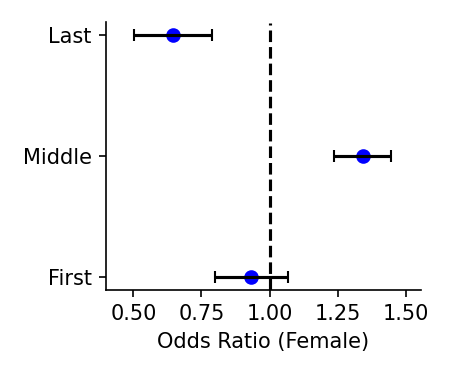}

\caption{Female to male odds ratio (95\% CI) of holding authorship position.  The vertical dashed line (i.e. Odds ratio = 1) represents no difference.}
\label{fig:Fig9}   
\end{figure}
Figure~\ref{fig:Fig9} represents the odds ratio ($OR$) of female to male at different authorship position. An odds ratio quantifies the strength of association between two events. The female to male odds ratio quantifies the likelihood of a female having an event over a male. An odds ratio gives a positive number greater than 0, where $OR == 1$ represents that females are as likely as males to hold a specific authorship position, whereas $OR < 1$ represents that females are less likely than males to hold a specific authorship position and vice-verse. As shown in figure~\ref{fig:Fig9}, it is evident that females are less likely than males to have both first $(OR = 0.93; [95\% CI : 0.67 - 1.19])$ and last $(OR = 0.65; [95\% CI : 0.36 - 0.92])$ authorship position in retracted papers whereas more likely to be the middle author $(OR = 0.1.34; [95\% CI : 1.13 - 1.54])$. 


\section{Conclusion}
The findings in this study sheds light on various facets of retractions in scholarly publications, particularly focusing on journal impact, retraction reasons, retraction time, country collaborations, and authorship patterns. The retraction timeline revealed significant peaks in retractions in the years 2021 and 2023. The analysis of journal impact highlighted that a substantial portion of retractions occurred in journals with high impact factors (Q1), emphasizing the need for stringent scrutiny and quality assurance mechanisms in prestigious publications. 

Reasons for retractions were multifaceted, ranging from authorship issues and conflicts to data-related concerns, journal-related issues, plagiarism, fake-biased peer review, unethical research, and unknown reasons. Notably, data-related concerns and multiple reasons emerged as predominant factors leading to retractions, underscoring the importance of upholding research integrity and transparency. The examination of time to retraction elucidated that a significant majority of retractions occurred within the first six months of publication, with variations observed across different journal quartiles, article types, and retraction reasons. This underscores the necessity for prompt detection and action to address research misconduct and errors. 

The analysis of country collaboration networks revealed intricate patterns of international and national collaborations among authors of retracted papers, with prominent collaborations observed between the USA, China, India, and other countries. Interestingly, while collaborations were prevalent, a considerable proportion of retractions stemmed from papers with national authorship, indicating potential challenges in research oversight and collaboration dynamics. Furthermore, the assessment of authorship patterns highlighted gender disparities in authorship positions, with females being less likely to hold first or last authorship positions but more likely to be middle authors in retracted papers. 

Overall, this study underscores the multifaceted nature of retractions in scholarly publications and emphasizes the importance of robust quality assurance measures, transparent reporting practices, and equitable authorship standards to uphold research integrity and foster a culture of responsible conduct in scientific inquiry. The results of the current study suggest that maintaining a high standard of ethics in professional endeavors is now more crucial than ever for scientists and researchers, as these difficulties will have a major impact on the reproducibility of research by authors and institutions. The journals, editors, reviewers, and gatekeepers of the caliber of scientific output must exercise extreme caution; otherwise, the research will be futile. Finally, it is primordial and noteworthy to remember that publishing fabricated, and falsified results, especially in health sciences, can damage patients' health and thus can cause irreparable damage to patients.


\section*{Data availability statement}
The data is freely available on Retraction Watch (\url{https://retractionwatch.com/retracted-coronavirus-covid-19-papers/}).

 \section*{Conflict of interest}
 The author declares no conflict of interest.
\printcredits
\bibliographystyle{cas-model2-names}

\bibliography{cas-refs}
\end{document}